\documentclass[aip,jcp,amsmath,amssymb,citeautoscript,reprint]{revtex4-2}
\usepackage[table]{xcolor}
\usepackage{colortbl}
\usepackage{graphicx, float, pgf}
\usepackage{physics, siunitx}
\usepackage[caption=false, justification=centerlast]{subfig}

\makeatletter
\let\oldtheequation\theequation
\renewcommand\tagform@[1]{\maketag@@@{\ignorespaces#1\unskip\@@italiccorr}}
\renewcommand\theequation{(\oldtheequation)}
\makeatother

\usepackage{hyperref}
\hypersetup{
    pdfauthor={Amartya Bose},
    pdftitle={Zero-Cost Corrections to Influence Functional Coefficients from Bath Response Functions},
    colorlinks=true,
    linkcolor=black,
    citecolor=black,
    urlcolor=black
}

\newcolumntype{e}{>{\columncolor{red!50}}c}

\begin{document}
\title{Zero-Cost Corrections to Influence Functional Coefficients from Bath Response Functions}
\author{Amartya Bose}
\affiliation{Department of Chemistry, Princeton University, Princeton, New Jersey 08544}
\allowdisplaybreaks

\begin{abstract}
    Recent work has shown that it is possible to circumvent the calculation of
    the spectral density and directly calculate the coefficients of the discretized
    influence functionals using data from classical trajectory simulations.
    However, the accuracy of this procedure depends on the validity of the high
    temperature approximation. In this work, an alternative derivation based on
    the Kubo formalism is provided. This enables the calculation of additional
    correction terms that increases the range of applicability of the
    procedure to lower temperatures. Because it is based on the
    Kubo-transformed correlation function, this approach enables the direct use
    of correlation functions obtained from methods like ring-polymer molecular
    dynamics and centroid molecular dynamics in determining the influence
    functional coefficients for subsequent system-solvent simulations. The
    accuracy of the original procedure and the corrected procedure is
    investigated across a range of parameters. It is interesting that the
    correction term comes at zero additional cost. Furthermore, it is possible
    to improve upon the correction using zero-cost physical intuition and
    heuristics making the method even more accurate. 
\end{abstract}
\maketitle

\section{Introduction}
Simulation of quantum dynamics in the condensed phase is a challenging problem.
Classical mechanics can often be a very approachable approach to such
simulations. These classical calculations miss out on corrections coming from
quantum dispersion and zero-point energy effects. A lot of work has been done
to incorporate quantum effects in classical trajectories ranging from full
semiclassical dynamics~\cite{vanvleckCorrespondencePrincipleStatistical1928,
hermanSemiclasicalJustificationUse1984, millerAlternateDerivationHerman2002} to
single classical trajectory-based approaches like the Wigner
approach~\cite{wignerRemarksTheoryReaction1939,
wignerQuantumCorrectionThermodynamic1932}, centroid molecular dynamics
(CMD)~\cite{caoFormulationQuantumStatistical1994Equilibrium,
caoFormulationQuantumStatistical1994Dynamical} and ring polymer molecular
dynamics (RPMD)~\cite{craigQuantumStatisticsClassical2004}. However, for
problems where the quantum nature of the dynamics is inevitable, these
classical trajectory-based approaches are not useful. Often in such cases, a
system-solvent decomposition can be performed limiting the quantum nature of
the dynamics to a low-dimensional subspace. Typically reduced density matrix
approaches related to the hierarchical equations of motion
(HEOM)~\cite{tanimuraTimeEvolutionQuantum1989} and quasi-adiabatic propagator
path integral (QuAPI)~\cite{makriTensorPropagatorIterativeI1995,
makriTensorPropagatorIterativeII1995} are used for such problems.

Recently tensor network approaches have been used in conjunction with both
HEOM~\cite{shiEfficientPropagationHierarchical2018, yanNewMethodImprove2020,
yanEfficientPropagationHierarchical2021} and
QuAPI~\cite{strathearnEfficientNonMarkovianQuantum2018,
boseTensorNetworkRepresentation2021, bosePairwiseConnectedTensor2022}. Based on
the tensor network representation of path integral using the Feynman-Vernon
influence functional~\cite{feynmanTheoryGeneralQuantum1963}, one can develop a
multi-site method that is capable of simulating extended quantum
systems~\cite{boseMultisiteDecompositionTensor2022}. This new multi-site tensor
network path integral has been used to study the dynamics and absorption
spectra of the B850 ring~\cite{boseTensorNetworkPath2022}. The presence of the
solvent makes the time propagation of system non-Markovian. In the path
integral framework, this non-Markovian memory is expressed as two-point
interactions, $\eta_{kk'}$, characterized by the separation between them. These
$\eta_{kk'}$ coefficients are related to integrals of the bath response
function, and have historically been expressed as integrals over the spectral
density. While this is convenient for model studies with analytical spectral
densities, it necessitates high quality molecular dynamics (MD) simulations for
estimating the spectral density when the solvent is atomistically given. The
presence of numerical noise in these molecular dynamics simulations along with
the requirement to simulate up to long times to reach equilibrium complicate
the calculation of these spectral densities.

Recently, Allen, Walters and Makri~\cite{allenDirectComputationInfluence2016}
have proposed a technique to directly use the energy gap autocorrelation
function to estimate the $\eta_{kk'}$ coefficients and avoid the computation of
the spectral density. While this classical approximation (CA) method is
extremely simple, its basic assumptions limit its applicability only to high
temperatures. This stems from the simultaneous identification of the classical
correlation function and its derivative with the real and imaginary parts of
the quantum correlation function respectively. It is quite well-understood that
the quantum dispersion and zero-point effects would affect the real part
significantly in all but the highest temperatures. The basic goal of this paper
is to derive a similarly computationally efficient method of obtaining the
discretized influence functional coefficients from the bath response function
but with better accuracy at lower temperatures.

Before going further, it would be prudent to note that this discussion is based
on the presumption that the potential energy surface describing the dynamics is
\emph{ab initio}. Though \emph{ab initio} molecular dynamics (AIMD) is becoming
increasingly approachable for large systems using neural network fits of the
forces and energies from density functional
theory~\cite{zhangDeepNeuralNetwork2020, zhangDeepPotentialMolecular2018},
classical force-fields such as
CHARMM~\cite{vanommeslaegheCHARMMGeneralForce2010} still remain extremely popular.
These classical force-fields are generally parameterized to include nuclear
quantum effects, thereby ensuring that the classical mechanics yields
experimentally correct results. When the solvent is described by such classical
force-fields that already account for nuclear quantum effects, the simultaneous
identification of the real and imaginary part of the quantum correlation
function with the classical correlation function and its derivative
respectively in CA is correct. The rest of the paper assumes some sort of an
\emph{ab initio} description that is not parameterized to account for the
nuclear quantum effects of the solvent.

In this paper, we propose an approach to deriving the discretized
influence functional coefficients directly in terms of the Kubo-transformed bath
response function. The Kubo-transform of the bath response function is given
by
\begin{align}
    \alpha_\text{Kubo}(t) &= \frac{1}{\beta Q}\int_0^\beta\dd{\lambda}\Trace\left[e^{-(\beta-\lambda)\hat{H}_\text{sol}}\,\hat{f}\,e^{-\lambda\hat{H}_\text{sol}}\hat{f}(t)\right].
\end{align}
The Kubo correlation function is commonly simulated by methods for simulating
approximate quantum dynamics like centroid molecular dynamics
(CMD)~\cite{caoFormulationQuantumStatistical1994Equilibrium,
caoFormulationQuantumStatistical1994Dynamical} and ring-polymer molecular
dynamics (RPMD)~\cite{craigQuantumStatisticsClassical2004}. While there are
many ways of expressing the quantum correlation function, the Kubo formulation
is widely known to be the most similar to the classical correlation function.
The explicit formulation of the $\eta$-coefficient in terms of the Kubo
correlation function presented here creates a clear link between methods like
CMD and RPMD and Feynman-Vernon influence functional-based path integral.

This new derivation not only provides an alternative to the CA
approach~\cite{allenDirectComputationInfluence2016}, but additionally gives a
full series expansion. Consequently it becomes easy to improve the results by
incorporating the leading order corrections, while retaining all the numerical
advantages of CA. We show that the most important first-order correction term
can be analytically simplified and written in terms of the bath response
function. The idea behind incorporation of these higher order terms is to
correct for the discrepancy brought in by assuming that the classical
correlation function does an adequate job of representing the real part of the
quantum correlation function. These higher order terms are dependent on
time-derivatives of the bath response functions. Numerical derivatives are
extremely sensitive to noise present in the data. So, we have expressed the
second-order correction in terms of a different correlation function. Every
extra order of correction either requires calculation of completely different
correlation functions or numerical derivatives. This makes it impractical to go
to very high orders. We have further improved the first-order correction
through physical arguments and heuristics. The errors in CA and the corrected
methods are evaluated with respect to the $\eta$-coefficients and the dynamics.
The most attractive aspect of the first-order and the heuristic corrections is
that they can be done completely free of any additional cost, retaining the
dependence only on the energy-gap correlation function.

The methods are derived in Sec.~\ref{sec:method} and a variety of numerical
tests and illustrations are shown in Sec.~\ref{sec:results}. Additionally, an
approach to rigorously obtain the next correction term is derived in the
Appendix. We end the paper in Sec.~\ref{sec:conclusion} with some conclusions
and observations regarding these efforts to use the classical or semiclassical
correlation function directly in the generation of the discretized influence
functional coefficients.

\section{Method}\label{sec:method}
Consider a quantum system coupled with a dissipative solvent:
\begin{align}
    \hat{H} &= \hat{H}_0 + \hat{H}_\text{sol},
\end{align}
where $\hat{H}_0$ is the Hamiltonian describing the quantum system and
$\hat{H}_\text{sol}$ describes the dissipative solvent. If the system is a
two-level system, $\hat{H}_0 = \epsilon\hat{\sigma}_z -
\hbar\Omega\hat{\sigma}_x$, where $\hat{\sigma}_{x,y,z}$ are the Pauli spin
matrices.

The thermal dissipative solvent in many cases is atomistically described. If
the fully atomistic description needs to be considered, one can use various
mixed quantum-classical methods such as the quantum-classical path integral
method~\cite{lambertQuantumclassicalPathIntegralI2012,
lambertQuantumclassicalPathIntegralII2012} for simulation. However, generally it
is possible to map the essentially anharmonic solvent onto a harmonic bath
under the Gaussian response theory. In such a case, the harmonic bath and its
interactions with the system is characterized by a spectral density,
\begin{align}
    J(\omega) &= \frac{\pi}{2}\sum_j\frac{c_j^2}{m_j\omega_j}\,\delta(\omega-\omega_j),
\end{align}
where $\omega_j$ and $c_j$ are the frequency and the coupling of the
$j$\textsuperscript{th} harmonic oscillator to the system. Under the harmonic
bath the Hamiltonian of the environment is given by:
\begin{align}
    \hat{H}_\text{sol} &= \sum_j \frac{\hat{p}^2}{2m_j} + \frac{1}{2} m\omega_j^2\left(\hat{x}_j - \frac{c_j \hat{s}}{m_j\omega_j^2}\right)^2,
\end{align}
where $\hat{s}$ is the operator that couples the system with the solvent.

The spectral density, for this harmonic mapping, is obtained from the solvent
energy gap correlation function also called the bath response function. It is
related to the spectrum corresponding to the bath response function as follows:
\begin{align}
    \alpha(\omega) &= \frac{2 J(\omega)}{1 - \exp(-\hbar\omega\beta)}.
\end{align}
Consequently, the bath response function is given by
\begin{align}
    \alpha(t) &= \frac{1}{\pi}\int_0^\infty\dd{\omega} J(\omega) \left(\coth\left(\frac{\hbar\omega\beta}{2}\right)\cos(\omega t) - i \sin(\omega t)\right).\label{eq:response_function}
\end{align}

If the initial condition is specified as a direct product of the system reduced
density matrix and the bath thermal density, then the system reduced density
matrix after $N$ time-steps of length $\Delta t$ can be represented as a path
integral,
\begin{widetext}
    \begin{align}
        \mel{s_N^+}{\rho(N\Delta t)}{s_N^-} & = \sum_{s_0^\pm}\sum_{s_1^\pm}\ldots\sum_{s_{N-1}^\pm} \bra{s_N^+}\hat{U}\dyad{s_{N-1}^+}\hat{U}\ket{s_{N-2}^+}\ldots\nonumber                      \\
                                            & \times\bra{s_1^+}\hat{U}\dyad{s_0^+}\rho(0)\dyad{s_0^-}\hat{U}^\dag\ket{s_1^-}\ldots\bra{s_{N-1}^-}\hat{U}\ket{s_N^-} F[\{s^\pm_j\}] \label{eq:pi} \\
        \text{where }F[\{s^\pm_j\}]         & = \exp\left(-\frac{1}{\hbar}\sum_{k=0}^{N}(s_k^+-s_k^-)\sum_{k'=0}^{k}(\eta_{kk'}s_{k'}^+ - \eta^*_{kk'}s_{k'}^-)\right).\label{eq:fvif}
    \end{align}
\end{widetext}
Here, $\hat{U}$ is the short time system propagator, $s_j^\pm$ is the
forward-backward state of the system at the $j$\textsuperscript{th} time point.
The Feynman-Vernon influence function~\cite{feynmanTheoryGeneralQuantum1963} is
denoted by $F[\{s^\pm_j\}]$, which is dependent upon the history of the path.
This influence functional can be described in terms of certain
$\eta$-coefficients~\cite{makriTensorPropagatorIterativeI1995,
makriTensorPropagatorIterativeII1995}, and can be expressed as double integrals
of $\alpha(t)$. The most general form is given as:
\begin{align}
    \eta_{kk'} &= \int_{(k-\frac{1}{2})\Delta t}^{(k+\frac{1}{2})\Delta t} \dd{t'}\int_{(k'-\frac{1}{2})\Delta t}^{(k'+\frac{1}{2})\Delta t} \dd{t''}\,\alpha(t'-t'').
\end{align}
The spectral density and the correlation functions required here are quantum
mechanical. However, owing to the large dimensionality of the solvent, one has
to resort to classical trajectory-based approximations to the quantum dynamics.
The most popular such approaches are
CMD~\cite{caoFormulationQuantumStatistical1994Dynamical} and
RPMD~\cite{craigQuantumStatisticsClassical2004}. These approaches estimate the
Kubo transform of a given correlation function
\begin{align}
    \alpha_\text{Kubo}(t) &= \frac{1}{\beta Q}\int_0^\beta\dd{\lambda}\Trace\left[e^{-(\beta-\lambda)\hat{H}_\text{sol}}\,\hat{f}\,e^{-\lambda\hat{H}_\text{sol}}\hat{f}(t)\right]
\end{align}
The main allure behind the Kubo-transformed correlation function is that it has
many similarities in structure with the classical correlation function. Thus,
departing from the classical approximation
(CA)~\cite{allenDirectComputationInfluence2016}, we assume that if only a
classical correlation function is available, it is more prudent to use it as an
approximation to the Kubo-transformed correlation function. (The rest of the
section deals only with $\alpha_\text{Kubo}(t)$ and its derivatives.)

It is well-known that the Kubo-transformed correlation function has identical
information to the standard correlation function. In particular, the standard
spectrum is related to the Kubo spectrum by
\begin{align}
    \alpha(\omega) &= \frac{\hbar\,\omega\beta}{1 - \exp(-\hbar\,\omega\beta)} \alpha_\text{Kubo}(\omega).
\end{align}
Because the Kubo correlation function is even and consequently, the Kubo
spectrum is symmetric, one can relate the spectral density to the Kubo
spectrum,
\begin{align}
    J(\omega) &= \frac{\hbar\,\omega\beta}{2} \alpha_\text{Kubo}(\omega).\label{eq:KuboJ}
\end{align}
It is possible to use the spectral density from the approximate quantum
calculations in our estimation of the $\eta$-coefficients. However, generating
accurate, noise-free quantum correlation functions for the Fourier transform
involved in the calculation of the spectral density can be challenging for
large systems. It has been shown that expressing the $\eta$-coefficients
directly in terms of the correlation function can make the approach numerically
robust~\cite{allenDirectComputationInfluence2016}.

We can relate the bath response function to the Kubo correlation function
estimated by substituting Eq.~\ref{eq:KuboJ} in Eq.~\ref{eq:response_function},
expanding the $\coth$ term to a series and doing the integrals
\begin{align}
    \alpha(t) &= \alpha_\text{Kubo}(t) - \frac{i\,\hbar\,\beta}{2}\dot\alpha_\text{Kubo}(t) + \frac{1}{3}\left(\frac{i\,\hbar\,\beta}{2}\right)^2\,\ddot{\alpha}_\text{Kubo}(t)\nonumber\\
              &- \frac{1}{45}\left(\frac{i\,\hbar\,\beta}{2}\right)^4\,\dv[4]{t}\alpha_\text{Kubo}(t) + \mathcal{O}(\hbar^6)\label{eq:standard_kubo}.
\end{align}
It is interesting that the series Eq.~\ref{eq:standard_kubo} has only a single
imaginary term. All terms other than
$\frac{i\hbar\beta}{2}\dot\alpha_\text{Kubo}(t)$ are real. This is only true
for the Kubo-transformed correlation function. There has been a lot of work
done on relating a classical correlation function to the corresponding quantum
correlation function~\cite{kimEvaluationQuantumCorrelation2002,
kimEvaluationQuantumCorrelation2006, egorovQuantumDynamicsVibrational1999}.
Such expansions structurally look similar to Eq.~\ref{eq:standard_kubo}.
However, when the quantum correlation function is expanded in terms of the
classical correlation function, there are higher order corrections to the
imaginary part as well~\cite{kimEvaluationQuantumCorrelation2002}.
Additionally, the harmonic approach to obtaining the quantum correlation
function from the classical correlation function has exactly the same form as
here. However, in the harmonic approach, it is an approximation, whereas here
it is rigorously true. 

If one uses a classical correlation function to approximate
$\alpha_\text{Kubo}(t)$ and includes only upto the term linear in $\hbar$ in
Eq.~\ref{eq:standard_kubo}, one would recover the results of
Ref.~\cite{allenDirectComputationInfluence2016}. To summarize, Allen, Walters
and Makri~\cite{allenDirectComputationInfluence2016} proposed that the real
part of the $\eta$-coefficients be obtained by doing a quadrature,
\begin{align}
    \Re\eta^{(0)}_{kk'} &= \int_{(k-\frac{1}{2})\Delta t}^{(k+\frac{1}{2})\Delta t} \dd{t'}\int_{(k'-\frac{1}{2})\Delta t}^{(k'+\frac{1}{2})\Delta t} \dd{t''}\,\alpha_\text{Kubo}(t'-t''),
\end{align}
(the superscript (0) is there to indicate that this is the uncorrected version)
and for the imaginary part,
\begin{align}
    \Im\eta_{kk'} &= \frac{\hbar\,\beta}{2}\int_{(k-\frac{1}{2})\Delta t}^{(k+\frac{1}{2})\Delta t} \dd{t'}\int_{(k'-\frac{1}{2})\Delta t}^{(k'+\frac{1}{2})\Delta t} \dd{t''}\,\dot\alpha_\text{Kubo}(t'-t''),
\end{align}
they evaluated the ``inner'' integral analytically, thereby transforming the
term into a single integral of the correlation function.

It is easy to see that the first order of correction to the real part of the
$\eta$-coefficients can be calculated analytically from
Eq.~\ref{eq:standard_kubo}. On analytically simplifying the expressions for the
first order corrections, one finds that it is in form of different linear
combinations of the values of Kubo correlation function
$\alpha_\text{Kubo}(t)$.
\begin{widetext}
    \begin{align}
        \Re\eta_{kk'} &= \Re\eta^{(0)}_{kk'} - \frac{\hbar^2\,\beta^2}{12}\int_{(k-\frac{1}{2})\Delta t}^{(k+\frac{1}{2})\Delta t} \dd{t'}\int_{(k'-\frac{1}{2})\Delta t}^{(k'+\frac{1}{2})\Delta t} \dd{t''}\,\ddot\alpha_\text{Kubo}(t'-t'')\\
                      &= \Re\eta^{(0)}_{kk'} - \frac{\hbar^2\,\beta^2}{12}\left(\alpha_\text{Kubo}((k-k'+1)\Delta t) - 2\alpha_\text{Kubo}((k-k')\Delta t) + \alpha_\text{Kubo}((k-k'-1)\Delta t)\right)\label{eq:truncated_kubo_first}\\
        \Re\eta_{00} &= \Re\eta^{(0)}_{00} - \frac{\hbar^2\,\beta^2}{12}\left(\alpha_\text{Kubo}\left(\frac{\Delta t}{2}\right) - \alpha_\text{Kubo}(0)\right)\\
        \Re\eta_{kk} &= \Re\eta^{(0)}_{kk} - \frac{\hbar^2\,\beta^2}{12}\left(\alpha_\text{Kubo}(\Delta t) - \alpha_\text{Kubo}(0)\right)\\
        \Re\eta_{k0} &= \Re\eta^{(0)}_{k0} - \frac{\hbar^2\,\beta^2}{12}\left(\alpha_\text{Kubo}\left(\left(k+\frac{1}{2}\right)\Delta t\right) - \alpha_\text{Kubo}(k\Delta t) + \alpha_\text{Kubo}((k-1)\Delta t) - \alpha_\text{Kubo}\left(\left(k-\frac{1}{2}\right)\Delta t\right)\right)\\
        \Re\eta_{N0} &= \Re\eta^{(0)}_{N0} - \frac{\hbar^2\,\beta^2}{12}\left(\alpha_\text{Kubo}(N\Delta t) - 2\alpha_\text{Kubo}\left(\left(N-\frac{1}{2}\right)\Delta t\right) + \alpha_\text{Kubo}((N-1)\Delta t)\right)\label{eq:truncated_kubo_last}
    \end{align}
\end{widetext}
Equations~\ref{eq:truncated_kubo_first}--\ref{eq:truncated_kubo_last} define
the current method that we would refer to as the first-order truncated Kubo
(TK1) approximation to the eta coefficients. Though the higher-order terms in
the series require the calculation of the numerical derivatives, it might be
possible to estimate them using different correlation functions. This is
derived explicitly for the second-order correction term in
Appendix~\ref{app:second-order}. This second-order truncated Kubo approximation
would be referred to as TK2.

The structure of the several approaches to approximating the quantum
correlation function in terms of the
classical~\cite{egorovQuantumDynamicsVibrational1999,
kimEvaluationQuantumCorrelation2002} are similar to the equations listed above.
The first correction term is indeed proportional to the second derivative of
the correlation function. The difference is only in the exact prefactor used,
all of which are of the form $\frac{\hbar^2\beta^2}{c}$. (Apart from $c=12$
derived here, $c=8$ also appears in certain approximations.) So, the correction
terms would also look very similar. 

While TK2 can be calculated if required, TK1 is by far the simpler of the two
algorithms. It needs no extra information than the bath response function. Let
us, therefore, analyze the TK1 approximation and see if we can use heuristics
to improve it. To motivate the changes, consider the behavior of the
$\eta$-coefficients as estimated by TK1 on lowering the temperature. First,
notice that the true quantum correlation function would become invariant to
temperature below a certain value. This is because as the temperature is
lowered, the thermal density matrix would asymptotically become the same as the
density matrix corresponding to the ground state. Consequently, the quantum
correlation function would asymptotically tend to the ground state correlation
function. However, this is not the case with the truncated Kubo approximations.

The uncorrected real part, $\Re\eta^{(0)}$, would show a behavior identical to
the Kubo correlation function. To understand the dependence on $\beta$,
consider the Kubo-transformed position autocorrelation function of a harmonic
oscillator,
\begin{align}
    C^{xx}_\text{Kubo}(t) &= \frac{1}{\beta m \omega^2} \cos(\omega t).
\end{align}
Clearly, the correlation function goes to zero as $\beta^{-1}$. Consequently,
$\Re\eta^{(0)}$ would also go to zero as $\beta^{-1}$. The correction term has
a prefactor $~\beta^2$, and hence overall would increase linearly with $\beta$.
This means that the corrected $\Re\eta$ terms would overall increase as
$\beta$. (This problem arises because we are truncating the infinite series in
Eq.~\ref{eq:standard_kubo} after the quadratic term in $\hbar$. In fact the
second-order term incorporated in TK2 is a worse offender. It would grow as
$\beta^3$ as $\beta\to\infty$. We will demonstrate in Sec.~\ref{sec:results}
that though TK2 increases the range of temperature where we can get good
$\eta$-coefficients, its errors increase extremely fast once out of this ``good
region.'') The imaginary part is independent of $\beta$ because of its
prefactor.

\begin{table*}
    \begin{tabular}{l|c|ce|ce|ce|ce|ce}
            & from $J(\omega)$ & CA & CA $\Delta$(\%) & TK1 & TK1 $\Delta$(\%) & TK2 & TK2 $\Delta$(\%) & tanh & tanh $\Delta$(\%) & tanh2 & tanh2 $\Delta$(\%)\\\hline
        $\Re\eta_{00}$   & $0.02932$     & $0.02921$     & $0.36928$     & $0.02932$     & $0.00162$     & $0.02932$     & $0.00003$     & $0.02932$     & $0.00093$    & $0.02932$     & $0.00113$    \\
        $\Re\eta_{10}$   & $0.11474$  & $0.11436$  & $0.33202$  & $0.11474$  & $0.00119$  & $0.11474$  & $0.00001$  & $0.11474$  & $0.00057$ & $0.11474$  & $0.00128$ \\
        $\Re\eta_{11}$   & $0.11675$     & $0.11633$     & $0.36154$     & $0.11675$     & $0.00153$     & $0.11675$     & $0.00002$     & $0.11675$     & $0.00085$    & $0.11675$     & $0.00117$    \\
        $\Re\eta_{20}$   & $0.10565$  & $0.10543$  & $0.20712$  & $0.10565$  & $0.00003$  & $0.10565$  & $0.00001$  & $0.10565$  & $0.00036$ & $0.10565$  & $0.00151$ \\
        $\Re\eta_{21}$   & $0.22566$  & $0.22497$  & $0.30538$  & $0.22566$  & $0.00092$  & $0.22566$  & $0.00001$  & $0.22566$  & $0.00035$ & $0.22566$  & $0.00136$ \\
        $\Re\eta_{30}$   & $0.09246$  & $0.09241$  & $0.05257$  & $0.09246$  & $0.00079$  & $0.09246$  & $0.00001$  & $0.09246$  & $0.00089$ & $0.09246$  & $0.00118$ \\
        $\Re\eta_{31}$   & $0.20500$  & $0.20465$  & $0.16938$  & $0.20500$  & $0.00021$  & $0.20500$  & $0.00001$  & $0.20500$  & $0.00053$ & $0.20500$  & $0.00147$ \\
        $\Re\eta_{20,1}$ & $0.01712$ & $0.01714$ & $0.07452$ & $0.01712$ & $0.00003$ & $0.01712$ & $0.00000$ & $0.01712$ & $0.00017$& $0.01712$ & $0.00059$\\
        $\Re\eta_{30,1}$ & $0.00767$ & $0.00767$ & $0.03531$ & $0.00767$ & $0.00001$ & $0.00767$ & $0.00000$ & $0.00767$ & $0.00007$& $0.00767$ & $0.00027$
    \end{tabular}
    \caption{Comparison of the methods for $s=1$, $\hbar\omega_c\beta = 0.15$, $\xi = 1$, $\Delta t = 1.25\hbar\beta$. Parameters taken from Ref.~\cite{allenDirectComputationInfluence2016}. Only the magnitudes of the relative percentage errors are reported. For this parameter there is practically no difference between the truncated Kubo and the tanh approaches, and both are practically exact. Where the differences are slightly more prominent (eg. the first three rows), the tanh correction seems to improve the results.}\label{tab:ohmic}
\end{table*}

\begin{table*}
    \begin{tabular}{l|c|ce|ce|ce|ce|ce}
            & from $J(\omega)$ & CA & CA $\Delta$(\%) & TK1 & TK1 $\Delta$(\%) & TK2 & TK2 $\Delta$(\%) & tanh & tanh $\Delta$(\%) & tanh2 & tanh2 $\Delta$(\%)\\\hline
        $\Re\eta_{00}$   & $0.02944$     & $0.02912$     & $1.09347$     & $0.02945$     & $0.00796$     & $0.02944$     & $0.00017$     & $0.02945$     & $0.00590$    & $0.02944$     & $0.00023$    \\
        $\Re\eta_{10}$   & $0.11029$  & $0.10926$  & $0.94097$  & $0.11030$  & $0.00530$  & $0.11029$  & $0.00008$  & $0.11030$  & $0.00353$ & $0.11029$  & $0.00173$ \\
        $\Re\eta_{11}$   & $0.11623$     & $0.11499$     & $1.06196$     & $0.11624$     & $0.00738$     & $0.11623$     & $0.00015$     & $0.11623$     & $0.00538$    & $0.11623$     & $0.00057$    \\
        $\Re\eta_{20}$   & $0.08515$  & $0.08482$  & $0.39150$  & $0.08515$  & $0.00219$  & $0.08515$  & $0.00012$  & $0.08515$  & $0.00292$ & $0.08515$  & $0.00508$ \\
        $\Re\eta_{21}$   & $0.20969$  & $0.20794$  & $0.83196$  & $0.20970$  & $0.00363$  & $0.20969$  & $0.00003$  & $0.20969$  & $0.00207$ & $0.20968$  & $0.00258$ \\
        $\Re\eta_{30}$   & $0.05368$  & $0.05392$  & $0.44952$  & $0.05368$  & $0.00814$  & $0.05368$  & $0.00013$  & $0.05368$  & $0.00729$ & $0.05368$  & $0.00474$ \\
        $\Re\eta_{31}$   & $0.15454$  & $0.15420$  & $0.21649$  & $0.15453$  & $0.00376$  & $0.15454$  & $0.00013$  & $0.15453$  & $0.00416$ & $0.15453$  & $0.00534$ \\
        $\Re\eta_{20,1}$ & $-0.01462$ & $-0.01462$ & $0.04407$ & $-0.01462$ & $0.00000$ & $-0.01462$ & $0.00000$ & $-0.01462$ & $0.00008$& $-0.01462$ & $0.00032$\\
        $\Re\eta_{30,1}$ & $-0.00717$ & $-0.00717$ & $0.02948$ & $-0.00717$ & $0.00000$ & $-0.00717$ & $0.00000$ & $-0.00717$ & $0.00006$& $-0.00717$ & $0.00022$
    \end{tabular}
    \caption{Comparison of the methods for a super-Ohmic bath with $s=2$. The rest of the parameters are identical to Table~\ref{tab:ohmic}. Even in the worst case scenario, the corrected methods (TK and the tanh correction) are two orders of magnitude better.}\label{tab:superohm}
\end{table*}

While to correct this issue, we would need to consider the infinite series,
here we give a poor man's \emph{ad hoc} approximate way of treating the
symptom. The uncorrected real part $\Re\eta^{(0)}$ is closely related to the
classical correlation function, so we do not change it. Now, turning to the
first order correction term. The current coefficient is
$-\frac{1}{3}\left(\frac{\hbar\beta}{2}\right)^2$. To get rough temperature
independence at very low temperatures, we need to have the prefactor grow
linearly with $\beta$ as that would compensate the $\beta^{-1}$ scaling of the
correlation function. However, this new term has to be equal to the current
coefficient, and grow as $\beta^2$ for $\beta\to 0$. The function $x\tanh(x)$
has this property of behaving like $x^2$ for small values of $x$ and as $x$ for
large values of $x$. So, we utilize this intuition to come up with two related
but slightly different splittings:
\begin{enumerate}
    \item Use $-\frac{1}{3}\left(\frac{\hbar^2\beta}{2E_h}\right)\tanh\left(\frac{E_h\beta}{2}\right)$ as the prefactor. This is called the tanh1 approximation.
    \item Use $-\frac{1}{3}\left(\frac{\hbar^2\beta}{4E_h}\right)\tanh\left(E_h\beta\right)$ as the prefactor. This is called the tanh2 approximation.
\end{enumerate}
(The Hartree energy terms are incorporated to make the argument of the
hyperbolic tangent function dimensionless. All the simulations here are done in
atomic units and consequently $\hbar=1$ and $E_h = 1$.) The logic behind
differentiating between these two different techniques for taking care of the
low temperature behavior is that in the first case, $\frac{\beta}{2}$ is taken
as a unit because it came from $\frac{\omega\beta}{2}$ in the reciprocal space.
In the second case, we do not keep the entire numerical multiplier outside the
hyperbolic tangent. These are both correct at high temperatures, but would have
different ranges of validity at low temperatures. Furthermore, though not done
here, the same idea can quite simply be extended to correct TK2 as well. 

Finally, it is important to reiterate that any correction, including more terms
from the series, Eq.~\ref{eq:standard_kubo}, the TK approaches or the tanh
approaches, is only required when the dynamics is \emph{ab initio}. If a
force-field is parameterized to experimental observables, and accounts for
quantum effects, then the classical trajectory results does not need further
correction to incorporate quantum effects. This is particularly the case with
the well-known CHARMM molecular
force-field~\cite{vanommeslaegheCHARMMGeneralForce2010}. In fact, doing so
would lead to a double counting of quantum effects, and consequently incorrect
results~\cite{boseWignerDistributionAdiabatic2018}. For such cases, the
obtained classical correlation function should be used as the real part of the
quantum correlation function. The series in Eq.~\ref{eq:standard_kubo} should
be truncated after the term $\mathcal{O}(\hbar)$.

\section{Results}\label{sec:results}
Consider the family of sub-Ohmic, super-Ohmic and Ohmic spectral densities with
exponential cutoffs, given generally as
\begin{align}
    J(\omega) &= \frac{\pi}{2}\hbar\xi\frac{\omega^s}{\omega_c^{s-1}}\exp\left(-\frac{\omega}{\omega_c}\right).\label{eq:spectral_density}
\end{align}
Because we have the spectral density, this model gives us a good testing ground
for exploring the accuracy of the various approximate approaches to calculating
the $\eta$-coefficients under a variety of situations. Here we are using the
classical correlation function as an approximation to the Kubo-transformed
correlation function. The classical correlation function can be obtained
analytically as the following integral
\begin{align}
    \alpha_\text{Kubo}(t) &= \frac{1}{\pi}\int_0^\infty\dd\omega\,\frac{2}{\hbar\omega\beta}\,J(\omega)\,\cos(\omega t)\\
                          &= \frac{\xi}{\beta}\,\frac{\cos\left(s\tan^{-1}(\omega_c t)\right)}{\sqrt{(1 + \omega_c^2\,t^2)^s}}\,\Gamma(s)\,\omega_c\label{eq:class_ohmic}
\end{align}
For the Ohmic bath ($s=1$), the classical correlation function, Eq.~\ref{eq:class_ohmic} reduces to the well-known Lorentzian form
\begin{align}
    \alpha_\text{Kubo}(t) &= \frac{\xi\omega_c}{\beta(1+\omega_c^2 t^2)}.
\end{align}

As a first comparison, let us consider high temperature parameters that where
the CA method would work the best. Following the original
work~\cite{allenDirectComputationInfluence2016}, we use an Ohmic bath with
$\hbar\omega_c\beta=0.15$, $\xi=1$, and $\Delta t=1.25\hbar\beta$. We compare
some representative values of $\Re\eta_{kk'}$. The imaginary values are not
reported because they are the same for all three methods. So, the errors would
be identical. The comparison is shown in Table~\ref{tab:ohmic}. While CA
performs quite well, the results obtained by the truncated Kubo and the tanh
methods are practically exact. There are no differences in the five places of
decimal reported in the table. The TK2 approach of course eliminates all error,
reducing the relative error to around $10^{-5}\%$, while the other methods have a
larger but still completely negligible error of around $10^{-3}\%$.

Keeping all the other parameters the same, let us change the Ohmic bath to a
super-Ohmic bath with $s=2$. The errors are shown in Table~\ref{tab:superohm}.
Unlike the Ohmic case, the errors in $\Re\eta_{kk'}$ when using CA are quite
significantly larger. The family of baths given by
Eq.~\ref{eq:spectral_density} reaches a maximum at $\omega = s\,\omega_c$. So,
for a super-Ohmic spectral density the maximum occurs at a higher frequency,
which is at a colder equivalent temperature making CA worse. The errors of
the TK and the tanh methods also grow, though they continue to remain less than a
hundredth of a percent, and possibly negligible from the standpoint of a
dynamics simulation. Unsurprisingly, TK2 performs the best here as well.

\begin{figure}
    \subfloat[Percentage error in the real part of the diagonal $\eta$-coefficients.]{\includegraphics{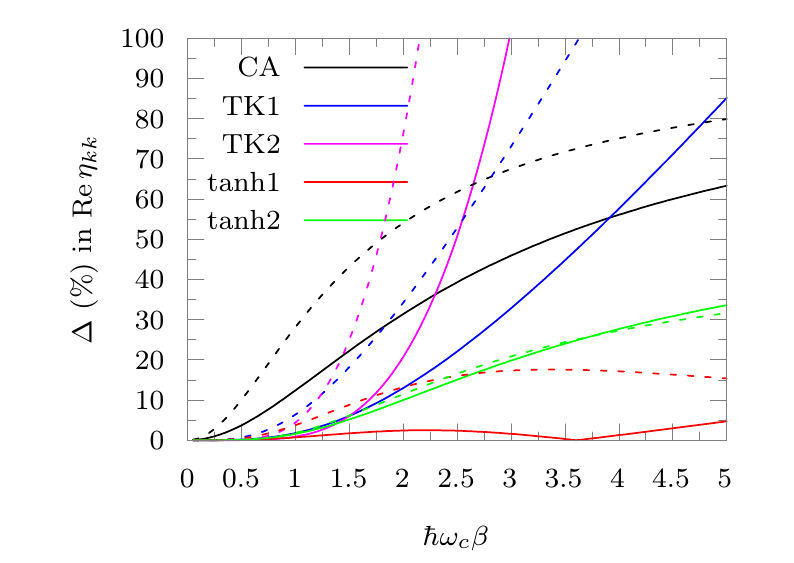}}

    \subfloat[Percentage error in the real part of the ``nearest neighbor interactions,'' $\eta_{k,k-1}$.]{\includegraphics{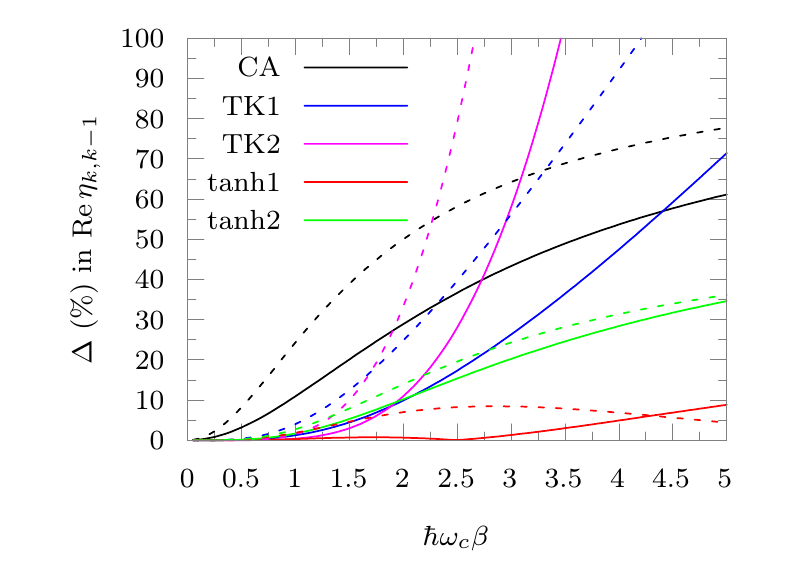}}
    \caption{Percentage error in various discretized influence functional coefficients for each of the methods. Solid lines: Ohmic spectral density. Dashed lines: Super-Ohmic spectral density ($s=2$). The plot goes to very cold temperatures as an illustration of the principle. The errors make the methods useless much earlier.}\label{fig:diagonal_error}
\end{figure}

\begin{figure*}
    \subfloat[Dynamics of the $\sigma_z(t)$.]{\includegraphics{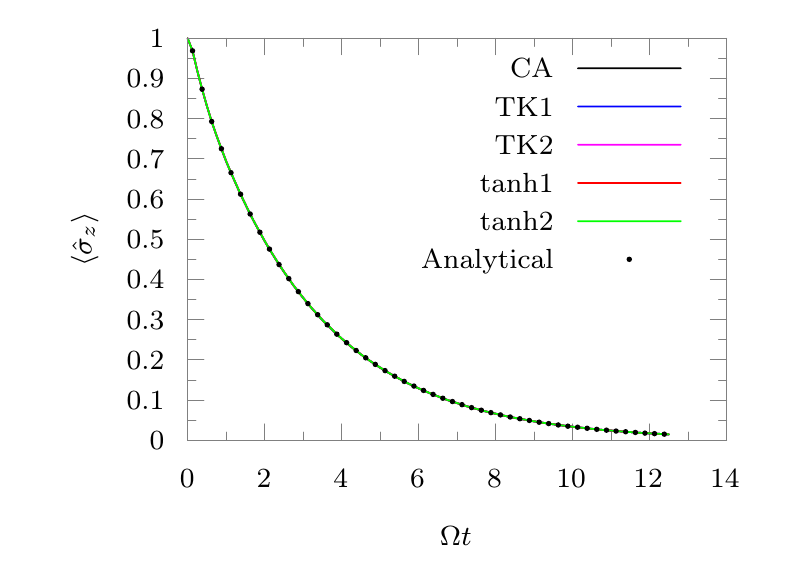}}
    ~\subfloat[Dynamics of the $\sigma_x(t)$.]{\includegraphics{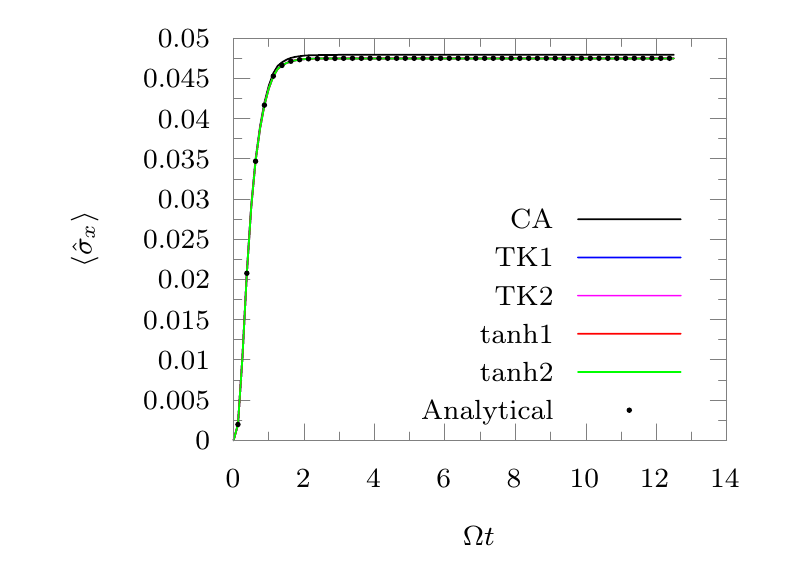}}
    \caption{Dynamics of various operators using the different methods of generating the $\eta$-coefficients at a temperature of $\hbar\omega_c\beta = 0.25$.}\label{fig:beta_2_5}
\end{figure*}

\begin{figure*}
    \subfloat[Dynamics of the $\sigma_z(t)$.]{\includegraphics{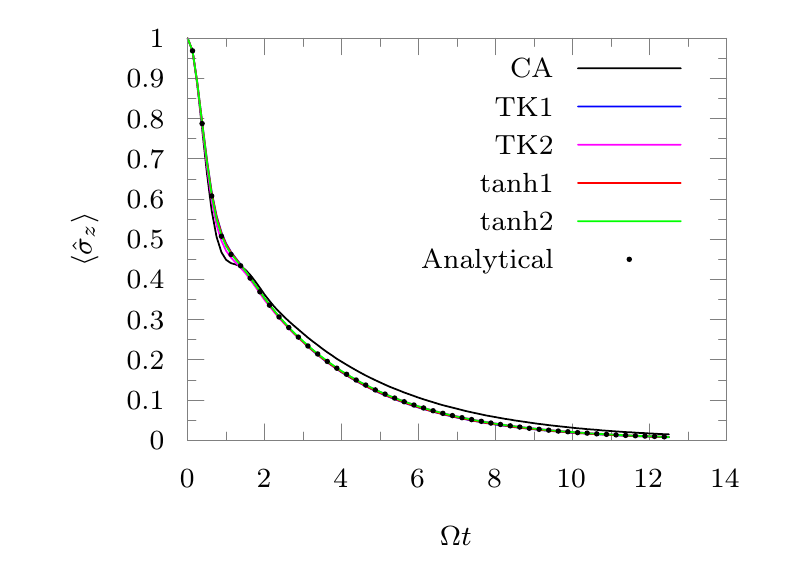}}
    ~\subfloat[Dynamics of the $\sigma_x(t)$.]{\includegraphics{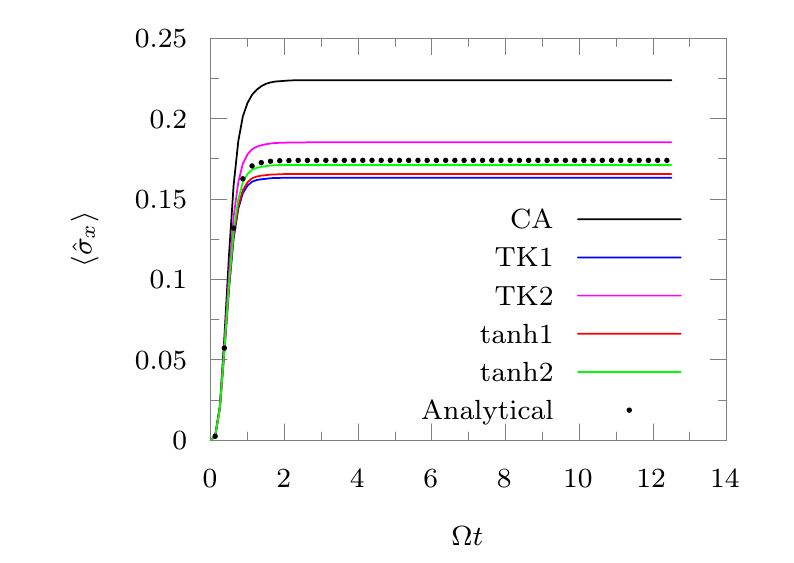}}
    \caption{Dynamics of various operators using the different methods of generating the $\eta$-coefficients at a temperature of $\hbar\omega_c\beta = 2$.}\label{fig:beta_2}
\end{figure*}

\begin{figure*}
    \subfloat[Dynamics of the $\sigma_z(t)$.]{\includegraphics{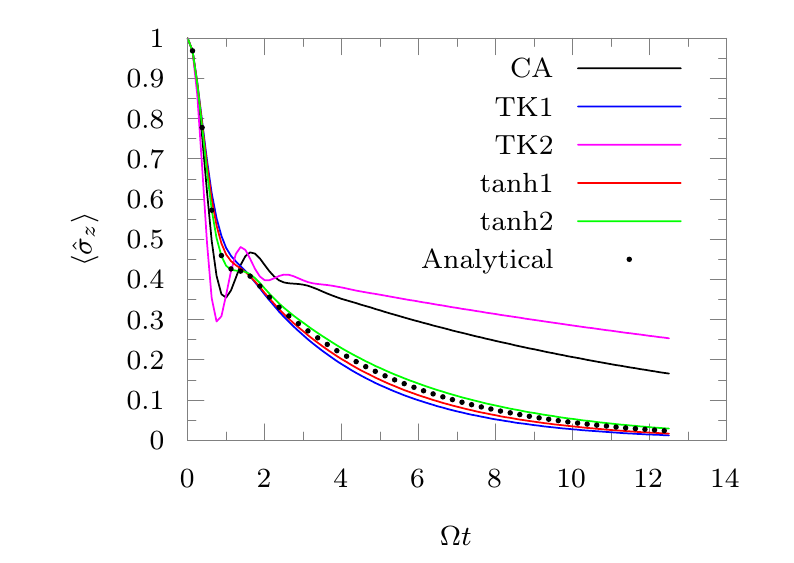}}
    ~\subfloat[Dynamics of the $\sigma_x(t)$.]{\includegraphics{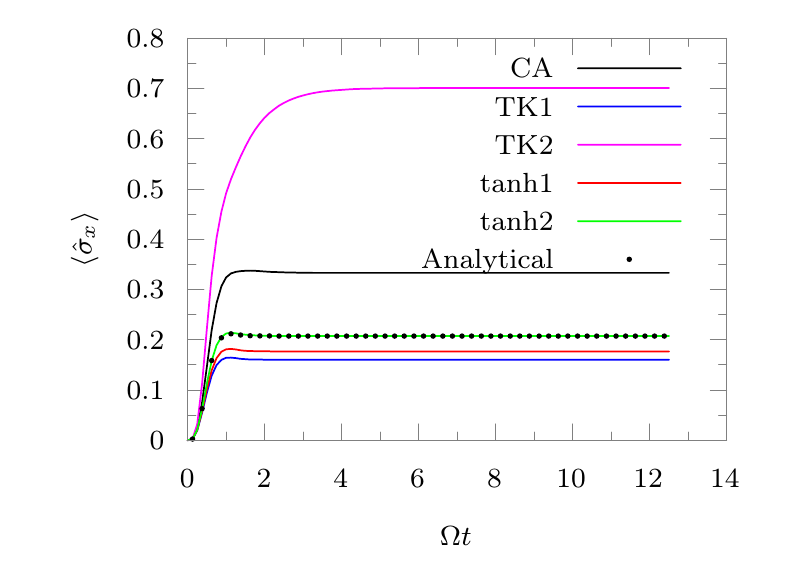}}
    \caption{Dynamics of various operators using the different methods of generating the $\eta$-coefficients at a temperature of $\hbar\omega_c\beta = 4$.}\label{fig:beta_4}
\end{figure*}

It is interesting to study the growth of the error in each of these methods
with increasing inverse temperature. Comparisons for the diagonal
$\Re\eta_{kk}$ terms and the terms connecting neighboring points,
$\Re\eta_{k,(k-1)}$ are shown in Fig.~\ref{fig:diagonal_error}. The range of
temperature illustrated in the figure is of course far beyond what any of the
three methods can handle. This is just a demonstration of how the errors grow
and not a statement about the usability of any of the methods. We can clearly
see that both of the corrected methods perform significantly better than the
original CA, increasing the region of applicability substantially. Using a 10\%
relative error as a threshold of applicability, we notice that CA is becomes
inaccurate at $\hbar\omega_c\beta = 1$. This is in comparison to the corrected
methods that extend the accuracy at least to $\hbar\omega_c\beta = 2$. Also, it
is extremely gratifying that our tanh correction hacks perform exceptionally
well, and consistently better than the base truncated Kubo method. In fact,
because of the discrepancy in the scaling of both TK1 and TK2 with $\beta$, we
find that though they do increase the range of applicability, beyond that the
error increases extremely fast. The fact that TK2 scales as $\beta^3$ while TK1
scales as $\beta$ for $\beta\to0$ gets reflected in the fact that TK1 becomes
more accurate beyond around $\hbar\omega_c\beta\approx 1.75$. Now, it is
debatable whether any of the TK methods should be used beyond that inverse
temperature any way.

While it is illuminating to explore the errors in certain $\eta$-coefficients,
it is at the end of the day not all that useful. The main problem is that it is
difficult to extrapolate errors in the $\eta$-coefficients to the error in the
dynamical observables one may be interested in. The discrepancies in certain
$\eta$-coefficients may not reflect as much and others may reflect more because
of the way the different path amplitudes interact. The only real way of judging
this is by simulating the dynamics of a TLS coupled to the bath by using the
different methods for calculating the influence functional coefficients.

For the examples with dynamics, we will consider a more difficult case for
these high temperature methods. The system is symmetric and defined by
$\hat{H}_0 = -\hbar\Omega\hat{\sigma}_x$. Consider an Ohmic bath with
$\xi=1.5$, $\omega_c=2.5\Omega$. The faster bath means that the effective
temperature would be lower. We start with a high temperature of
$\hbar\omega_c\beta = 0.25$. This is the regime where all the methods should be
equivalently good. The dynamics of $\hat\sigma_z$ and $\hat\sigma_x$ are shown
in Fig.~\ref{fig:beta_2_5}. While the dynamics of $\hat\sigma_z$ is identical
in all the methods, we can, even at this high temperature, visibly see the error
in the CA simulation of $\hat\sigma_x(t)$. Both the truncated Kubo and the tanh
corrections agree exactly with the result from the analytic
$\eta$-coefficients. Therefore, this error cannot be a result of any error in
$\Im\eta_{kk'}$ because that is common to all three methods. Lowering the
temperature to $\hbar\omega_c\beta=2$, in Fig.~\ref{fig:beta_2}, one finds that
CA method has fallen apart. Both TK and the tanh corrections continues to give
acceptable results for $\hat\sigma_z(t)$ but the TK1 and TK2 results for
$\hat\sigma_x(t)$ does not match the analytical results. While the tanh1
approach seems to be quite close to the TK1 method, tanh2 continues to give
extremely good agreement with the fully analytical result.

In fact, even at $\hbar\omega_c\beta=4$, the tanh2 correction gives quite
acceptable results. This is shown in Fig.~\ref{fig:beta_4}. At this very low
temperature, we see the effects of the extreme sharp rise of errors in
the TK2 method. We had mentioned that though TK2 increases the range of
applicability of the method, at low temperatures the scaling of TK2 ($\beta^3$)
is very different from the theoretical limits, where the correlation function
should be independent of $\beta$. Here we see such a very low temperature,
where the TK2 approach is in fact even worse than CA.

Thus, we see that all the methods discussed here significantly increase the
range of applicability of the correlation function-based approach to
calculating the discretized influence functional coefficients. Through rigorous
derivations of TK1 and TK2, we have increased the range of temperatures by
at least two-fold. It is pleasantly surprising that with tanh2, we have
attained an almost four-fold increase in the temperature range over which we
can get accurate dynamics.

\section{Conclusion}\label{sec:conclusion}
We have developed a systematic way for expressing the $\eta$-coefficients in
terms of the Kubo-transformed bath response function. This makes it possible to
use the results from methods like RPMD and CMD to characterize the harmonic
mapping of an \emph{ab initio} atomistic solvent onto a bath of harmonic
oscillators. The Kubo-transformed correlation function, though identical in
information content to standard quantum correlation function, is more classical
in the symmetries that it has. Therefore, if a classical correlation function
is used as an approximation to the Kubo-transformed correlation function and
this expression is truncated at the first order in $\hbar$, it reduces to the
classical approximation~\cite{allenDirectComputationInfluence2016}. While the
classical approximation was an \emph{ad hoc} approach, the relationship derived
over here can be used to rigorously converge the values of the
$\eta$-coefficients.

In addition to the general series, we have presented a host of useful and cheap
corrections to the classical approximation, thereby increasing the temperature
range over which one can directly compute the discretized influence functional
coefficients from correlation functions. We have shown how the first major
correction to the classical approximation scheme can be analytically integrated
out leading to a very simple change to the exact expressions. This change can
be obtained for no extra computational cost over the integrals required to get
the classical approximation. This first change is called the first-order
truncated Kubo expression. Though further terms can, in principle, be
incorporated, they require calculation of numerical derivatives or separate
costly correlation functions. If the original data is noisy, such computations
are often numerically unstable. Thus it is tempting to stop at this first
order. Interestingly, for the case of the second-order correction, we have
derived a relationship of this term with a different correlation function that
can also be estimated quite efficiently by molecular dynamics.

Because TK2 requires calculation of an extra correlation function, it is not
always very lucrative. It is interesting to think about improving TK1 with some
heuristics. We have analyzed the behavior of TK1, and shown that the real part
scales linearly with the inverse temperature, $\beta$. This is a problem
because at very low temperatures, the correlation function should
asymptotically tend to the ground state correlation function. The cause was
seen to be the truncation of the infinite series. However, we have proposed a
``poor-man's'' approximation that involves transforming the coefficient of the
correction term to respect the proper limits. This constitutes the motivation
behind the two tanh approximations that have been derived.

We have numerically assessed the performance of the various approximations
introduced here to the classical approximation and the analytical
$\eta$-coefficients obtained using the expressions in
Ref.~\cite{makriTensorPropagatorIterativeI1995}. (These numerically
explorations have been done on a harmonic bath, where the classical and Kubo
correlation functions are the same. For calculations on anharmonic solvents,
ideally an approximate Kubo correlation function like one from RPMD or CMD
should be used. In absence of such an approximation, the classical correlation
function can also be used because of the similarities that the Kubo function
shares with it.) We showed that the exact values of the $\eta$-coefficients
calculated by the TK and the tanh approaches, even at high temperatures, are
significantly closer to the analytical results compared to the CA method. The
newly introduced corrections yield more accurate $\eta$-coefficients through
the entire applicable range of temperature. Despite this overall increase of
accuracy, it is noted that TK2 breaks down quite pathologically. This has been
understood from the perspective of scaling with the inverse temperature,
$\beta$, which can only be truly solved by considering the entire infinite
series or, equivalently, working in the frequency domain. While the true bath
response function becomes independent of $\beta$ at very low temperatures, the
TK approximations do not. In fact the TK2 approach grows as $\beta^3$. This is
what leads to the pathological breakdown. Because the various tanh
approximations were built to fix this problem heuristically, it is very
encouraging that they perform significantly better than TK approaches at low
temperatures.

The values of these discretized influence functional coefficients, while
extremely crucial, interact with the path amplitude and the path sum in
non-trivial ways making an extrapolation of errors in coefficients to errors in
dynamics impossible. We simulated the dynamics of a two-level system coupled to
an Ohmic bath at different temperatures with coefficients being derived by the
each of the four approximate correlation function-based approaches. The
corrections introduced here increase the temperature range of applicability of
the correlation function-based approach by almost four times in the best case.
Of course, if the simulation is done at a cold enough temperature and all the
higher order terms are necessary, it would possibly be the simplest to
calculate the spectral density directly because of problems with the numerical
derivatives and the scaling with $\beta$.

The incorporation of Kubo correlation functions in path integral-based
approaches to system-solvent quantum dynamics seems a very lucrative way of
including anharmonic nuclear quantum effects in the solvent in a simple way.
Here we have just scratched the surface of this deep relationship. Future work
would look into further connections and possibilities.

\section*{Acknowledgments}
I acknowledge the support of the Computational Chemical Science Center:
Chemistry in Solution and at Interfaces funded by the US Department of Energy
under Award No. DE-SC0019394.

\appendix
\section{Relation between the second-order correction and classical correlation function}\label{app:second-order}
In the body of the text, we have focused on the first-order correction to CA,
primarily because that is the most important term and can be evaluated at zero
additional cost. While the higher-order corrections can be generally obtained
using appropriate numerical derivatives, it is possible to obtain the
second-order correction quite efficiently using a different set of correlation
functions.

Consider the second-order correction term to the bath response function,
Eq.~\ref{eq:standard_kubo}, given as $-\frac{\hbar^4\beta^4}{720}\dv[4]{t}
\alpha_\text{Kubo}(t)$. The corresponding correction to the discretized
influence functional coefficients would be given as a double integral of the
same:
\begin{align}
    \varepsilon_{kk'} &= -\frac{\hbar^4\beta^4}{720}\,\int_{(k-\frac{1}{2})\Delta t}^{(k+\frac{1}{2})\Delta t} \dd{t'}\int_{(k'-\frac{1}{2})\Delta t}^{(k'+\frac{1}{2})\Delta t} \dd{t''}\,\alpha^\text{(iv)}_\text{Kubo}(t'-t'').
\end{align}
By analytically doing the integrals, we can express the correction as
\begin{align}
    \varepsilon_{kk'} &= -\frac{\hbar^4\beta^4}{720}\left(\ddot{\alpha}_\text{Kubo}((k-k'+1)\Delta t)\right.\nonumber\\
                      &- \left.2\ddot{\alpha}_\text{Kubo}((k-k')\Delta t) + \ddot{\alpha}_\text{Kubo}((k-k'-1)\Delta t)\right)\label{eq:second_corr}
\end{align}
So, the second-order correction term to CA requires second-order derivatives of
the correlation function.

For simplicity let us assume that the classical correlation function is used
instead of the Kubo-transformed correlation function. So, we replace
$\alpha_\text{Kubo}$ with $\alpha_\text{Cl}$. Since RPMD and CMD both are
classical trajectory-based methods for approximating the correlation function,
a similar derivation can also be done for the relevant expressions
corresponding to the two approximately quantum methods. The bath response
function in its quantum form and its classical approximation are given as
\begin{align}
    \alpha(t) &\propto \expval{\hat{f}(t)\hat{f}(0)}\\
    \alpha_\text{Cl}(t) &\propto \iint\dd{\vb{q}}_0\dd{\vb{p}}_0 e^{-\beta \mathcal{H}(\vb{q}_0, \vb{p}_0)}\,f(\vb{q}_0)\,f(\vb{q}_t).
\end{align}

The first derivative of $\alpha_\text{Cl}(t)$ can be expressed as
\begin{align}
    \dot\alpha_\text{Cl}(t) &\propto \iint\dd{\vb{q}}_0\dd{\vb{p}}_0 e^{-\beta \mathcal{H}(\vb{q}_0, \vb{p}_0)}\,f(\vb{q}_0)\left(\vec{f'}(\vb{q}_t)\cdot\frac{\vb{p}_t}{m}\right).
\end{align}
Here, $\vec{f'}$ is the gradient of the function $f$. The second temporal
derivative can be expressed by the chain-rule
\begin{align}
    \ddot\alpha_\text{Cl}(t) &\propto \iint\dd{\vb{q}}_0\dd{\vb{p}}_0 e^{-\beta \mathcal{H}(\vb{q}_0, \vb{p}_0)}\,f(\vb{q}_0)\left(\frac{\vb{p}_t^\text{T}}{m}\cdot\overline{\overline{f''}}(\vb{q}_t)\cdot\frac{\vb{p}_t}{m}\right.\nonumber\\
                             &-\left.\vec{f'}(\vb{q}_t)\cdot\frac{\vb{F}(\vb{q}_t)}{m}\right)\label{eq:dderiv}.
\end{align}
The second-order derivative of $f$ with respect to the position is denoted by
$\overline{\overline{f''}}(\vb{q}_t)$. Generally storing such second-order
derivatives is challenging, but here, one can calculate them on-the-fly and
directly calculate the dot products. The force on the particle is given by
$\vb{F}$.

Thus it is possible to express the second-order time derivative of the
autocorrelation function in terms of cross-correlation functions,
Eq.~\ref{eq:dderiv}. Consequently, if required, one can evaluate correction
term, Eq.~\ref{eq:second_corr} in terms of this cross-correlation function.

\bibliography{bibexport.bib}
\end{document}